\documentclass[conference]{IEEEtran}
\usepackage[latin9]{inputenc}
\usepackage{array}
\usepackage{url}
\usepackage{multirow}
\usepackage{amsmath}
\usepackage{graphicx}

\makeatletter

\providecommand{\tabularnewline}{\\}

\makeatother

\usepackage[overlay,absolute]{textpos}

\begin{document}

\begin{textblock*}{10in}(23.5mm, 10mm)
{\textbf{Ref:} \emph{International Joint Conference on Neural Networks (IJCNN)}, pages 1--7, Rio de Janeiro, Brazil, July 2018.}
\end{textblock*}

\title{\textbf{DeepOrigin: End-to-End Deep Learning for Detection of New Malware
Families}}

\author{\IEEEauthorblockN{Ilay Cordonsky} \IEEEauthorblockA{Deep Instinct Ltd\\
ilayc@deepinstinct.com} \and \IEEEauthorblockN{Ishai Rosenberg} \IEEEauthorblockA{Deep Instinct Ltd\\
ishair@deepinstinct.com} \and \IEEEauthorblockN{Guillaume Sicard} \IEEEauthorblockA{Deep Instinct Ltd\\
guillaumes@deepinstinct.com}\and\IEEEauthorblockN{Eli (Omid) David}\IEEEauthorblockA{Deep Instinct Ltd\\
david@deepinstinct.com}}

\maketitle

\begin{abstract}
In this paper, we present a novel method of differentiating known from previously unseen malware families. We utilize transfer learning by learning compact file representations that are used for a new classification task between previously seen malware families and novel ones. The learned file representations are composed of static and dynamic features of malware and are invariant to small modifications that do not change their malicious functionality. Using an extensive dataset that consists of thousands of variants of malicious files, we were able to achieve 97.7\% accuracy when classifying between seen and unseen malware families.
Our method provides an important focalizing tool for cybersecurity researchers and greatly improves the overall ability to adapt to the fast-moving pace of the current threat landscape.
\end{abstract}

\IEEEpeerreviewmaketitle{}

\section{Introduction}

For the past decade deep learning advancements
influenced problem solving methods in various fields like natural
language processing and computer vision. Previously, to create a proper
model in each of those fields, one had to use advanced domain specific
techniques in order to extract human engineered features, which later
were used as heuristics or features in machine learning models. Deep
learning introduced representation learning, which made those tasks
much easier and today some tasks produce better results than humans
\cite{LeCun2015}.

While deep learning has been revolutionizing some fields, others still have
room to grow: The cyber space deals with 250,000 new threats daily\footnote{\url{https://www.av-test.org/en/statistics/malware/}},
however the traditional way of mitigating and analyzing new threats
heavily relies on human intervention. Clients will get infected,
upon noticing, they will propagate the incidents to specialized
teams, which in turn will investigate the malicious file. If a file
was already seen in the wild, the infection might be removed. However,
if it is a new variant or family, the team will have to investigate the
file by analyzing some aspects of its behavior (network activity, system calls, etc.).
As a security researcher, it is critical to quickly understand whether
one is dealing with a new type of malware, as the human resources in the
cyber defense area is very scarce and it is crucial for them to focus on the newest threat families. Doing so
heavily relies on domain expertise, and with an exponential growth
of malware we witness everyday, the task of differentiating between
known and previously unseen malware families becomes much harder.
Metamorphic and polymorphic malwares are rewritten numerous times such
that every iteration is different from the others. This type of obfuscation
makes it almost impossible to be detected by a standard signature based
anti-virus. Online forensics tools such as VirusTotal\footnote{http://www.virustotal.com}
will provide the security analyst with historical data of the file
only if the specific file was already found in the wild. In practice, many new variants
are undocumented and thrive unnoticed. 

To tackle the issue of differentiating known malware from the unknown we leverage 
automatic file signature generation. Moreover, the process
should be invariant to the file's static nature and its actual behavior,
both of which can be meddled with, thus obfuscating the initial static
or dynamic structure of the file. While some file signature generation
methods have already been proposed \cite{David2017}, none of them
were used to differentiate previously unseen malware from obfuscated
versions of known malware. The question we are trying to answer
is: Can we generate a low dimensional malware signature that captures
the file's main functionality in ways that allow for differentiation
between known and unknown malware families?

Our main contributions in this paper are:
\begin{itemize}
\item A novel method for file signature generation. We use these signatures
to construct a new classifier that distinguishes between known malware
families and new ones, unseen in the wild ones. The initial input generation
is automatic and does not rely on domain-specific knowledge
or any specific aspect of the file (static or dynamic). Primarily we are
using a supervised multi-class classifier (trained on already seen
labeled malware) to generate low dimensional file signatures that
in turn are used as input to a new classifier that solves the issue
of measuring functional differences between malware families. 

\item A unique threat landscape visualization method that allows for rationalizing about malware novelty: Unseen malware generates low neuron outputs. Plotting these outputs on a low dimensional space gathers new malware relatively close to the origin point and scatters known malware at a distance. This overview is a powerful forensics tool that also provides a glimpse into the neural network.
\end{itemize}

The proposed method consists of four stages:

\begin{enumerate}
\item A deep neural network is trained for known malware family classification.
\item Non-linear dimensionality reduction model is created by removing the
softmax layer from the deep neural network. The output of this model is the low
dimensional file representation. 
\item File representations are created for a second training dataset. This
dataset consists both of known and novel malware families.
\item A second classifier\footnote{The new classifier is evaluated using a test set that consists of distinct malware families that are not available during training.} is trained using output from the previous stage. This classifier learns to distinguish between known and novel malware families.
\end{enumerate}
We use an extensive dataset of 14 malware families that we split between the train and test sets. The split is temporal, i.e., to allow us to follow each family representation over time, all samples in the test set are from a later time than samples in the train set. Furthermore, we use another set of novel ransomware families, to differentiate them from the representations of the known families. The quality of the signatures is assured by analyzing the test accuracy of the original classifier, which achieves 97\% accuracy. To be able to reason about the signature space and visualize it, another model is trained using only two neurons for the ($n-1$)th layer. Removing the last layer leaves us with a 2D signature generator. Finally, we scatter plot the resulting signatures.

The rest of the paper is structured as follows: Section 2 summarizes
previous related work, and Section 3 provides the details of the proposed methodology.
In Section 4 we present experimental results, and Section 5 contains our concluding remarks and proposals for future work.

\section{Background and Related Work}

Nowadays multiple cybersecurity vendors already provide machine learning based malware detection solutions. State-of-the-art detection mechanisms usually offer a binary decision, i.e., whether the file is malicious or not. Some even tackle the malware family classification problem, but even then, the classification only occurs between a small number of known
malware families. Even the best malware classifier will not be able to provide an answer to a question that all malware researchers asks themselves with each new incident: did we already see the likes of this sample?

In practice, while such models might generalize well enough to catch a zero-day malware, further investigation of this file will be similar to that of any other malicious file detected by the same model, and no specific measures will be taken. Had we had an indicator that would expose samples belonging to a previously unseen new malware family, the incident handling might be completely different (e.g., calling a cybersecurity researcher to reverse engineer the malware and to evaluate both potential damage and possible removal techniques).

We will review some of the methods that previously addressed similar tasks. Comar et al.\cite{comar2013combining} created a 2-tier system where the first tier differentiates between malicious and benign files, while the second tier dives deeper into malware classification, discriminating between known malware and obfuscated variants of it. Using one-class SVM, they assembled hyperspheres within the feature space. Each malware family class is assigned to a separate hypersphere, and samples that do not fall in all the spheres are considered obfuscated versions of known malware. Training such a system suffers from scalability issues since one will have to train as many classifiers as the number of known malware families to establish a commercial grade system. Moreover, each time a new family is detected, a new model has to be trained for that specific family. Additionally, only network related data were extracted for the feature space, which makes the task domain-specific. In contrast, our method is using a deep neural network to utilize multiple levels of non-linearity between the features (as opposed to linear kernel-based SVMs, that only achieve a single level of non-linearity). Moreover, we encompass both dynamic and static characteristics of the file within our initial input space, e.g., API calls, registry entry modifications, network inspection, import/export tables, strings, etc. Thus, in addition to improved accuracy due to the usage of more features, we are not bound to any specific maliciousness indicator and therefore able to detect families of non-network related malware. A further difference is in how we define unseen malware: In our case, obfuscated versions of the same malware family should reside in proximity on the decision plane. Malware family of a distinct type that incorporates different behavior and intent will be considered as new, instead of recognizing zero-day as a polymorphic version of a known malware type. The rationale is that those samples have the same behavior as their variants, and therefore, there is no need to reverse engineer them again, unlike the novel malware families. 

PAYL \cite{wang2004anomalous} uses unsupervised learning, creating profiles of byte frequencies distributions per host and port connected to an application. They then decide whether the behavior is anomalous by testing if the Mahalanobis distance of the operational profile from the precomputed one is higher than a threshold set during the training period. Thus, this method is domain specific and uses only features extracted from the network activity of each application.

Deep learning is already being used for malware classification successfully. Dahl et al. \cite{dahl2013large} presented a successful malware classifier that uses random projections for initial dimensionality reduction, and a neural network classifier to distinguish between malicious families and benign files. However, their method does not address the task of differentiating between previously seen and unseen malware families.

David et al. \cite{David2017} analyzed each file in a Cuckoo sandbox environment where the files are executed in a secure environment and their behavior recorded (network activity, system calls, etc.). Each resulting report was tokenized and converted to a binary vector, which was converter later using stacked denoising autoencoders to a low dimensional signature. This method is different from our approach to signature generation, where we are using the output of the last hidden layer of a DNN classifier as our signatures. The current paper also provides an additional continuous metric for how novel a malware sample is, instead of just a binary value of whether the malware belongs to a known malware family or not. 

\section{Proposed Methodology}

In this section, we present our approach for file signature generation, for the task of differentiating between malware of known and unknown families. Additionally, we introduce a method to visualize the signatures in a low dimensional space. This visualization method is an important tool that allows for improved malware analysis.

The challenge we are dealing with is a classification task, in the sense that we would like the representations of different files from the same malware family to reside close to each other in the signature space, allowing for a clear separation between them and a different, unknown malware family set. Thus, the representation should be fuzzy enough to disregard slight modifications in the malware that only affect the structure of the file, without any behavioral change (e.g., appending a byte to the end of the file, changing the file header in a way that won't compromise its validity, modify the source code without changing the functionality, etc.). We hypothesize that if we use the pre-softmax layer output as files' signatures\footnote{To generate a signature for a new file, we use the trained network without the softmax layer - a file vector is fed to a forward pass of the network. The resulting output vector is the signature.}, their properties would allow separating between new and old malware. That is, for family A of new malware samples, we will get signature vectors of low floating point values, in contrast to samples from known malware family B that will produce signature vectors of higher values. By measuring the relative distance of each signature from the origin point (in the signature dimension space), we will be able to distinguish novel malware from known malware families, as the former will be closer to the origin point. By setting a threshold on this distance, we will assign a decision boundary for the question at hand. We can also use the distance from the origin to evaluate the ``novelty level'' of the malware.

\begin{figure}
\begin{centering}
\includegraphics[width=0.9\columnwidth]{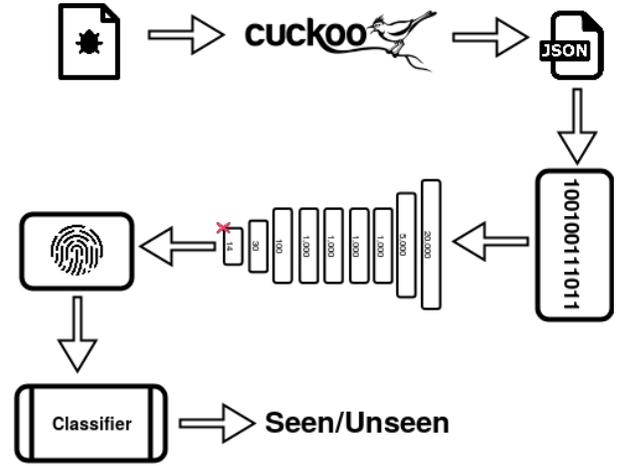}
\par\end{centering}
\caption{Simplified illustration of the proposed method \label{fig:Simplified-illustration-of}}

\end{figure}

The preprocessing phase, as shown in Figure \ref{fig:Simplified-illustration-of} consists of executing each malware in a sandbox framework, which in turn produces a text file in JSON format with various aspects of the file (i.e., API calls, network activity, static characteristics such as strings, etc.). 
Each of these perspectives may reveal a different angle to the scanned file's maliciousness. For example, if a file creates a new readme file, using the \texttt{CreateFile} API, and then encrypts some files with the \texttt{BCryptEncrypt} API, it may be ransomware. Plain text strings recovered from an executable may reveal the imported functions used, some of which may be alarming. High entropy values of various file sections may indicate that the file is packed\footnote{A packed executable is encoded when inactive and decoded at runtime directly into memory. Packing is a known technique used for obfuscation to avoid static analysis tools.}. Analyzing the network activity may expose downloads of unwanted artifacts, a method popular among malicious dropper software. The opening of ports to create backdoors is another network exploitation technique.

The JSON file is tokenized and converted to a high dimensional binary vector that is used as an input to the DNN classifier. The classifier consists of 9 fully connected layers, including a 30-neuron pre-softmax layer. The DNN uses PReLU activation function, and the last layer is a 14-neuron softmax layer since we are training it to differentiate between 14 known malware families. The softmax layer is then removed, allowing us to use the resulting model as a low dimensional signature generator.

\subsection{Initial File Vectorization}

Executables can be analyzed both statically and dynamically. The static representation of the file consists of all features that can be extracted without executing the file. Examples of such features are plain strings, and various file header parameters (e.g., compilation time, architecture, etc.). To extract a dynamic representation, we are required to run the file in a secluded environment (sandbox), and record the actions the file is performing (e.g., network packets sent and received, open sockets,
IO operations, API calls along with their arguments, etc.). Static features have two main advantages: 1) Faster and safe extraction (no need to run the file), and 2) Exploring all code flows of the analyzed file. However, dynamic features are harder to obfuscate. To combine the advantages of all feature types, we base our method on both.

Executing malicious files using Cuckoo Sandbox provides us with both static and dynamic features, and the output is stored in a structured JSON file. This file is the input to our preprocessing engine.

The absence of feature engineering allows the neural network to learn the most efficient feature representations. As a minimal preprocessing phase, we create a string dictionary. This dictionary is generated using the training data exclusively. We first tokenize each report to unigrams (words) and then choose the 20,000 most common unigrams as our dictionary. To eliminate zero variance features (i.e., JSON field names that exist in all reports, and hold no information within them), we remove all the words that are in all of the reports. We then perform a second pass on the data, to create a Boolean feature vector per Cuckoo output file. 
Each vector consists of binary bit-string, where each bit corresponds to a single unigram present or absent within the file. Those 20,000-dimensional vectors are used as input to the deep neural network described in the next subsection.

\subsection{Training the Deep Neural Network}

As stated before, our principal goal in this paper is to highlight new malware. We first create low dimensional file signatures, such that different variants of the same family get similar representations. We use these representations as input to a different classifier that will be trained upon them to differentiate between already known families and new ones. This method is somewhat similar to transfer learning \cite{pan2010survey}, a technique widely used for computer vision tasks.  For our purposes, we train a deep neural network classifier and use its last hidden layer output as input to a new classifier. We also introduce new malware families previously unseen by the original classifier. We use a simple linear classifier over the output values, and use each vector's Euclidean distance from the origin as a classification feature.

We train 9-layer DNN (20,000--5,000--1,000--1,000--1,000--1,000--100--30--14 neurons in each layer). 20,000-dimensional
binary vectors are used as input, and the output layer is 14-dimensional softmax layer. To achieve better generalization we perform batch normalization \cite{Ioffe2015} after each layer, along with dropout \cite{Srivastava2014} and input noise \cite{David2017} to increase input variance. For activation we chose PReLU \cite{He2015}. We use categorical cross-entropy as the loss function.

After training the model, we remove the 14-dimensional softmax layer, leaving a 30-dimensional pre-softmax layer as the new output layer. The values are not normalized and may vary across the positive and negative side of each axis. These floating point values serve as the low dimensional signature of the file. We test the hypothesis that for unknown malware family samples, most of the neurons will produce a low output value, considerably lower than that of known family samples.

\subsection{Visualization}

For visualization purposes, we train a separate neural network, where the pre-softmax layer consists of two neurons. Following training and removal of softmax layer, we use the output of these two neurons as a 2D file signature generator. These signatures are easily plotted and may be used as an additional tool for malware analysis.

\section{Experimental Results}

\subsection{Dataset}

For the baseline dataset, we chose to work with 14 different malware families: Nimnul, Dinwod, Delf, Blocker, Expiro, Kykymber, Zbot, Wabot, PornoAsset, Sytro, PolyRansom, Virut, WBNA, and Parite. The amount of families that we chose for the initial classification task is considered quite high. The rationale was to encompass a broad variety of malware behavior, so we created a dataset that consists of ransomware, banking
trojans, worms, backdoors, remote access, and remote administration tools.

The initial dataset consists of 9,922 files and is split between a train set of 7,759 and a test set of 2,163 samples. The split is performed by a complete temporal separation between the two sets, i.e., all samples in the train set appeared in the wild before 18-JAN-2017, while the ones in test set appeared after that date\footnote{According to VirusTotal ``first submitted'' value, a commonly-used value by security researchers}. Both training and test sets contain the same 14 malware families. Additionally, we created another dataset of 160 new ransomware samples from different families all of which were previously unseen within the initial dataset. These are used to evaluate the final classifier (the one that differentiates between seen and unseen malware).

\subsection{Previously Known Malware Families}

The initial model was trained using a dataset containing both old
and new malware families, spanning over the years 2012-2016. To familiarize the reader with the families at hand we provide short descriptions.

\textbf{Nimnul}: File infector that opens a backdoor. Using this
backdoor, an attacker can instruct the computer to download malicious content that can further damage your computer.

\textbf{Dinwod }: Trojan. Injects code to running processes and installs other malware.

\textbf{Delf }: Trojan. This is a generic code name. These trojans
are known for redirecting web traffic, application manipulation,
and installation of other malware.

\textbf{Blocker}: Generic ransomware code name. This type of malware encrypts sensitive data using an encryption key it
gets through the TOR network. It then presents a ransom message that asks for Bitcoin payment to decrypt files.

\textbf{Expiro}: Malware from the polymorphic file infectors family. It injects unique malicious code each time while preserving the malicious intent. The infected files are injected with information theft routines, which are executed when the original files are called.

\textbf{Kykymber}: Spyware that is usually dropped by a dropper malware. It meddles with registry, downloads more malicious files, and then deletes itself.

\textbf{Zbot}: (aka Zeus) Trojan. It steals personal and financial
information. May also lower browser security configuration and turn off the firewall.

\textbf{Wabot}: Opens a backdoor that allows for remote access.

\textbf{PornoAsset}: Trojan. Usually dropped by other malware droppers, or unknowingly downloaded when visiting malicious or compromised websites. Usually it will attack using a screen-locking notification that lists the demands of the malicious actors.

\textbf{Sytro}: Mass-mailing worm. This worm distributes by sending itself through email.

\textbf{PolyRansom}: Polymorphic file infector that infects the host with ransomware.

\textbf{Virut}: Virus. It infects executables, ASP, HTML and PHP
files. It spreads by replicating itself to removable and network drives.

\textbf{WBNA}: Worm. Spreads by replicating itself to removable and network drives. May also download and execute other malware.

\textbf{Parite}: Virus. It infects files on local file systems and on
network drives. The infection is by code injection. The infected file executes the malicious code, which in turn return the control to the original code, so there is no immediate indication of malicious behavior.

\subsection{Previously Unseen Malware Families}

For the differentiation task, we had to choose malware that appeared recently. Since in 2017 we witnessed a large number of ransomware attacks, it would be suitable to distinguish between ``old'' malware (that also includes some ransomware families) and new ransomware families that caused several billion dollars of damages throughout 2017. The following are some of the new families that we chose for the test set.

\textbf{Cerber}: Ransomware. First seen in mid-June 2016. It spreads via email attachment disguised as a Microsoft Word document. When the recipient opens the attached file, a macro executes automatically and downloads the Cerber ransomware.

\textbf{HydraCrypt}: Appeared in late 2016. It is distributed by exploit kits, and through URLs embedded in spam e-mails. Loading the URL will download the malware that in turn will encrypt data.

\textbf{WannaCry}: One of the most famous recent ransomware. Used in May 2017 to attack more than 230,000 machines in 150 Countries. Among the targeted services were hospitals, telecom companies, train companies and many more.

\textbf{Locky}: Released in 2016, highly active in 2017. Also delivered by email, as an artifact downloaded by a malicious
macro within a Microsoft Word document.

\textbf{NotPetya}: Together with WannaCry, one of the most damaging ransomware. It had massive success in infecting thousands of machines throughout 2017. One of its main differences from previously seen ransomware is that it spreads on its own (in contrast to the others that required it to be downloaded from a spam email). The most disturbing fact
about this malware is that although it acts like ransomware, it
actually is not. The file encryption is beyond repair, and paying
the ransom will not help with decrypting the data.

\subsection{Sandboxing}

All the files mentioned above are executed using Cuckoo\footnote{\url{https://cuckoosandbox.org/}} sandbox. Cuckoo is a popular file analysis tool which executes the
file in a secluded environment, and records various static and dynamic aspects of it (e.g., API calls along with arguments, network behavior, open ports, strings, registry manipulation, etc.). The results are saved in a structured JSON file and tokenized as presented in the previous section. After the tokenization, we project them to a 20,000-dimensional binary vector. These vectors serve as input to the deep neural network classifier.

\subsection{Training the DNN classifier}

\begin{figure}
\begin{centering}
\includegraphics[width=1\linewidth]{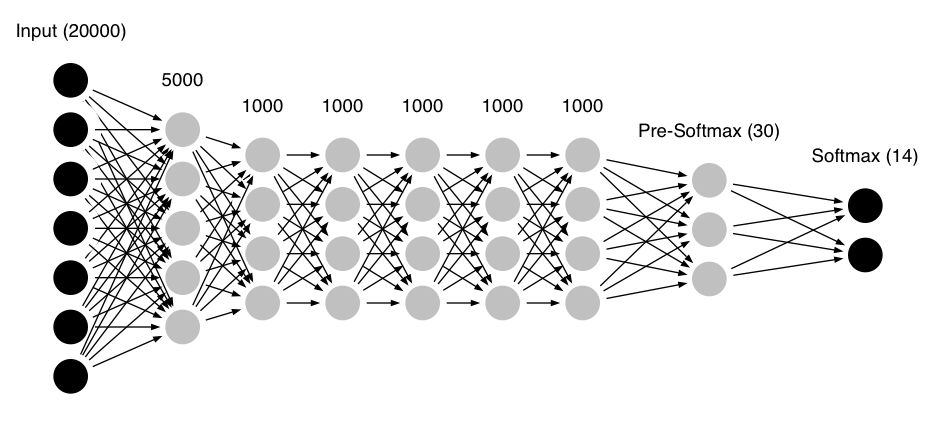}
\par\end{centering}
\caption{Proposed DNN diagram. All layers are separated by batch normalization and dropout. \label{fig:DNN-Visf}}
\end{figure}

To create low dimensional signatures, we trained a 9-layer DNN classifier. The network consists of 20,000 dimensional input layer, additional fully connected layers of 5,000, 1,000, 1,000, 1,000, 1,000, 100, 30 dimensions respectively, and a final 14-neurons softmax layer (Figure \ref{fig:DNN-Visf}). We applied batch normalization \cite{Ioffe2015} before each layer to deal with internal covariate shift. Internal covariate shift occurs where the distribution of inputs per layer changes following changes in the network parameters during the training phase. Applying batch normalization helps to gain faster convergence, and also adds another layer of regularization. We also applied dropout with the rate of 0.4, i.e., for each layer 40\% of randomly chosen neurons are turned off for each training sample, making neurons more independent of one another. Input noise was also applied at the same rate. While dropout is applied to reduce dependence between neurons, input noise is applied to achieve a higher variance of the input. Since we are projecting each Cuckoo report on a 20,000-dimensional vector, it is possible for two different files to have the same initial representation. By randomly turning off some of the indicators, we are making it almost impossible for the network to encounter the same input vectors.

When comparing the test set classification accuracy between PReLU and ReLU activated models, PReLU provides much-improved accuracy (97\% vs. 91.9\%). The difference between the two activation functions appears below. While ReLU cancels all negative feedback, PReLU allows it. Note that PReLU includes a slope argument $\alpha$ which is updated during training.

\begin{center}
${\displaystyle PReLU(\alpha,x)={\begin{cases}
\alpha x & {\text{for }}x<0\\
x & {\text{for }}x\geq0
\end{cases}}}$
\par\end{center}

\begin{center}
${\displaystyle ReLU(x)={\begin{cases}
0 & {\text{for }}x<0\\
x & {\text{for }}x\geq0
\end{cases}}}$
\par\end{center}

To summarize the network parameters: 10\% of training samples were used for validation. We applied dropout ratio 0.4 on each layer, including the input layer; We also applied batch normalization on each layer's input, using momentum of 0.99, centering and scaling, beta initialization of zeroes, gamma initialization of ones, moving mean initial values of ones, and moving initial variance values of ones. We used the PReLU
activation function, initializing alpha with zeros. For the last layer,
as stated before, we used a 14-dimensional softmax layer. The training ran for 1000 epochs, using a batch size of 32 samples with Adam optimizer \cite{Kingma2014}. Training was conducted on a single Nvidia GeForce GTX 1080 GPU. 

The classifier achieved 98.1\% validation accuracy and 97\% test accuracy. We then stripped the model of the softmax layer. Thus, each forward pass on the model would produce a 30-dimensional output vector of floating point values. These values are used as the low dimensional fingerprint of the malicious file. Note that we did not use benign files for this model since our goal is to differentiate between known and unknown families of malicious files (given the fact that the file is malicious).

\subsection{Visualization of New Malware}

Recapping our original assumption: for new malware families, the pre-softmax layer will not fire extreme output values, placing the 30-dimensional point closer to the origin, in comparison to known malware families. \footnote{Note that we're not dealing with benign files at all, the goal is, having a malicious file, assess its novelty. We're not interested in how benign files would appear in this signature space.}

\begin{figure*}
%\begin{centering}
\includegraphics[width=1\textwidth]{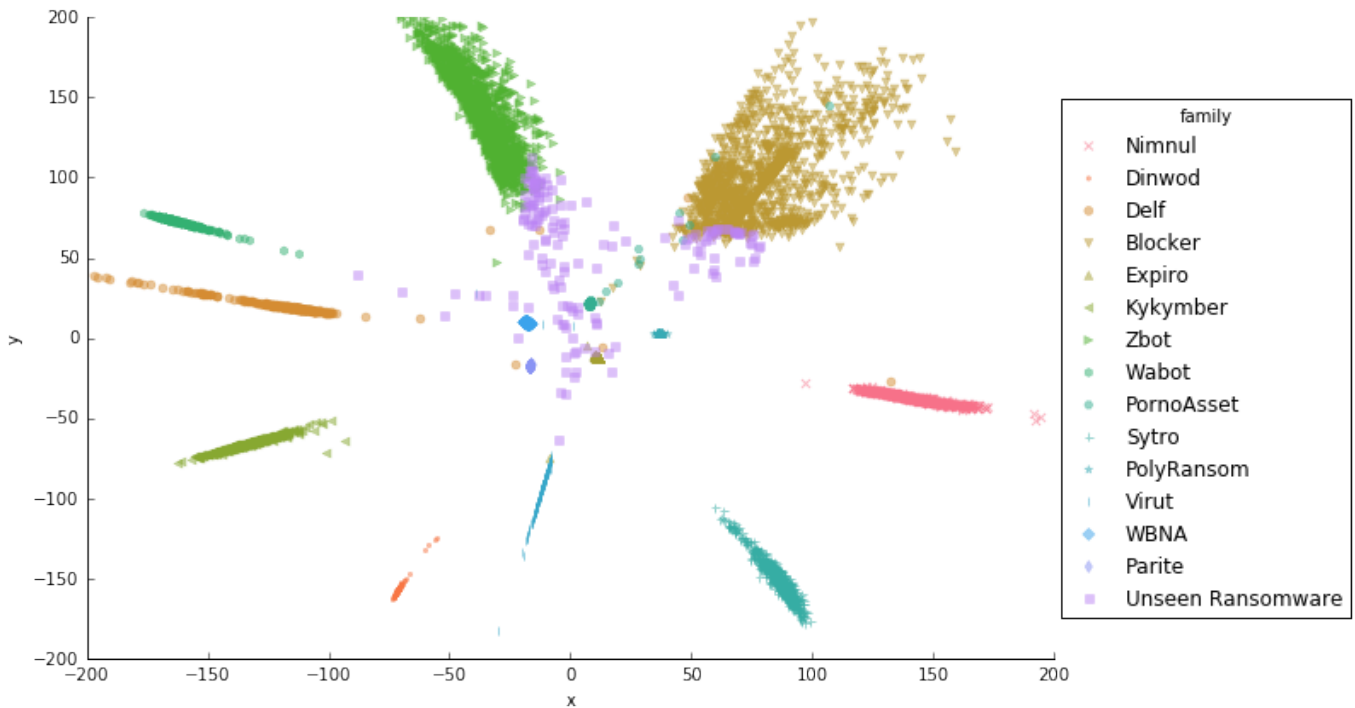}
%\par\end{centering}
\caption{Pre-softmax layer visualization of the 2D model version. Recent unseen
malware is represented by a rectangular marker. 
\label{fig:Pre-softmax-layer-visualization}}
\end{figure*}

Using dimensionality reduction methods such as t-SNE \cite{maaten2008visualizing} the distance between clusters is meaningless. Therefore, we trained another model, using a two-dimensional pre-softmax layer (two neurons in the last hidden layer). This layer's output can be easily plotted, an example of which is shown in Figure \ref{fig:Pre-softmax-layer-visualization}. While this is different from the original model (since changing a layer within a network will affect all the learned parameters), it does provide a visualization of how this layer affects the final classification decision, and how the model interprets known and unknown malware families.

Notice that unseen malware seems to group closer to the origin point than most the other clusters. We can still see other clusters close to the origin point, due to loss of information when thinning the pre-softmax layer from 30D to 2D. Another way to explain the proximity is by the similarity in behavior of the said malware due to, e.g., a mislabeled malware sample (since the current labeling process is manual). When measuring distributions of the 30D signatures, the unseen cluster appears to be by far the closest one. We show these measurements below.

\subsection{Analyzing Signature Space and Decision Boundary}

Visualizing the two-dimensional signature space (Figure \ref{fig:Pre-softmax-layer-visualization}) provides basic intuition to the pre-softmax layer behavior. This view is merely an approximation of the 30-dimensional signature space. To obtain a clearer picture we need to assess some of the clusters` probabilistic properties. Measuring mean and variance of each cluster will provide sufficient view on the pre-softmax output. As shown in Table \ref{tab:Distribution-of-the}, the distribution of the recent malware samples is considerably closer than the test set points of known families. The separation between the different groups allows for a classifier based on a certain threshold of the distance from the origin (i.e., if a file signature is closer to the origin than a certain threshold $T$, the classifier will infer this point as unseen malware).

\begin{table}
\begin{centering}
\begin{tabular}{|c|c|c|}
\hline 
\multirow{2}{*}{Family} & \multicolumn{2}{c|}{Distance from origin}\tabularnewline
\cline{2-3} 
 & mean & std.\tabularnewline
\hline 
\hline 
Blocker & 14.244988 & 2.674046\tabularnewline
\hline 
Delf & \multirow{1}{*}{14.978935 } & 2.001384\tabularnewline
\hline 
Dinwod & 27.014787  & 2.105340\tabularnewline
\hline 
Expiro & 20.268087 & 3.465324 \tabularnewline
\hline 
Kykymber & 23.500490 & 2.690603\tabularnewline
\hline 
Nimnul & 17.992709 & 2.705298\tabularnewline
\hline 
Parite  & 20.872712 & 4.088434\tabularnewline
\hline 
PolyRansom & 15.263254 & 4.904691\tabularnewline
\hline 
PornoAsset & 16.534225 & 3.367623\tabularnewline
\hline 
Sytro & 22.870228 & 0.980014\tabularnewline
\hline 
Virut & 15.750393 & 3.520776\tabularnewline
\hline 
WBNA & 20.325697 & 1.412213\tabularnewline
\hline 
Wabot & 23.067854 & 0.498773\tabularnewline
\hline 
Zbot & 14.457342 & 1.951387\tabularnewline
\hline 
\textbf{Recent Ransomware} & \textbf{7.920390} & \textbf{2.198623}\tabularnewline
\hline 
\end{tabular}
\par\end{centering}
~

\caption{Distribution of the different malware families over the 30D signature
space. \label{tab:Distribution-of-the} }
\end{table}

\begin{figure}
\begin{centering}
\includegraphics[width=0.7\linewidth]{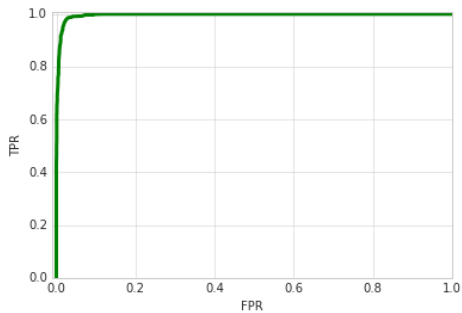}
\par\end{centering}
\caption{ROC curve of the final classifier that is used to differentiate between
known and unknown malware. \label{fig:ROC-curve-of}}
\end{figure}

Finally, we created a threshold-based classifier. To find the decision boundary (that is, the distance under which the malware is considered new), we used ROC statistics: Using the test
set samples, we plotted True Positive Rate (TPR) vs. False Positive
Rate (FPR), as shown in Figure \ref{fig:ROC-curve-of}. The calculated AUC was 0.996, and the best balance between TPR and FPR was obtained using a threshold of 10.2, which resulted in 96.3\% TPR, 2.1\% FPR, and accuracy of 97.7\%. The statistics were created using a new set of 3,800 malicious samples, evenly balanced between the seen and unseen malware families.

\section{Concluding Remarks}

In this paper, we presented a method for separating between known and unknown malware families. To capture each file's behavioral and static nature, we ran it in a sandbox environment and generated a textual report. The report was tokenized and later converted to a large sparse vector. We fed the resulting vectors (of known malware families) to a deep neural network. We then removed the pre-softmax layer to create a signature generation model that converted each sample to a 30-dimensional vector. We analyzed the output vector space to obtain a decision boundary based on the Euclidean distance of each point from the origin. This distance is a metric that measures the novelty of a malware family, i.e., its behavioral difference from existing families. 

To the best of our knowledge, this method was never used for such tasks within cybersecurity, and can be used as a practical
tool in a security researcher's arsenal; knowing that one is dealing
with new malware that was never seen in the wild usually means that extensive measures should be applied to this sample, since early recognition of new malware families substantially mitigates the threat. 

By detecting such new malware early, fewer machines will be infected, less private and sensitive information will be stolen, and less money will be lost due to ransom payments and other damages.

Our future work would consist of scaling our method by training the initial DNN using extended datasets of more known malware families. We will also explore unsupervised training approaches for this task.

\newpage 

\bibliographystyle{IEEETran}
\bibliography{bibtex}

\end{document}